\documentclass[12pt]{article}
\usepackage[hmargin=1in,vmargin=1in]{geometry}
\usepackage{setspace}
\usepackage{amsfonts,amsmath,amssymb,amsthm}
\usepackage{natbib}
\usepackage{lscape}
\usepackage{graphicx}
\usepackage{enumerate} 
\usepackage{booktabs}
\usepackage{adjustbox}
\usepackage[toc,title,titletoc,header]{appendix}
\usepackage{multirow}
\usepackage{threeparttable} 
\usepackage{xcolor} 
\usepackage{float} 
\usepackage{etoolbox} 
\usepackage[colorlinks,allcolors=blue!70!black]{hyperref}
\allowdisplaybreaks

\usepackage{tikz}
\usepackage{pgfplots}
\usepgfplotslibrary{groupplots} 
\pgfplotsset{compat=1.18}     

\usepackage{setspace}  
\usepackage{lipsum}
\usepackage{caption}
\usepackage{graphics}
 \usepackage{subfigure}
 \usepackage{graphicx}

\usepackage{bm}
\usepackage{bbm}
\usepackage{comment}         
\usepackage{multicol}
\usepackage{multirow}
\usepackage{caption}
\usepackage{braket}

\usepackage{etoolbox}

\usepackage{pgfplots}
\pgfplotsset{compat=1.18}

\newcommand\E {\mathbb{E}}

\usepackage{color}

\usepackage{graphicx}

\newcommand{\R}{\mathbb{R}}

\newcommand{\1}{\mathbbm{1}}

\usepackage{times}

\newcommand{\pa}{\mathrm{\pa}}

\newcommand{\RN}[1]{%
  \textup{\uppercase\expandafter{\romannumeral#1}}%
}

\usepackage{tabularx} 
\usepackage{booktabs} 
\usepackage{ragged2e} 
\usepackage{geometry} 

\title{
Distributional Treatment Effects of Content Promotion:  
Evidence from an ABEMA Field Experiment\footnote{
The authors are grateful to 
Editor of this special issue, Toru Kitagawa and Aleksey Tetenov, and an anonymous referee for their insightful comments and constructive suggestions, which have significantly improved the quality of this manuscript. We also thank the participants at the 2023 Conference on Digital Experimentation at MIT (CODE@MIT) for their helpful comments and suggestions.
Oka acknowledges the financial support
from JSPS KAKENHI Grant Number 24K04821. 
}
}

\author{ 
  \begin{tabular}[t]{cc}
    \begin{minipage}[t]{0.30\textwidth}
      \centering
      Shota Yasui\footnote{
      AI Lab, 
      CyberAgent 
      (Email: 
      \href{mailto:yasui_shota@cyberagent.co.jp}{yasui\textunderscore shota@cyberagent.co.jp}
      )
      } \\
    \end{minipage}
    &
    \begin{minipage}[t]{0.30\textwidth}
      \centering
      Tatsushi Oka\footnote{
      Department of Economics, 
      Keio University
      (Email: 
      \href{mailto:tatsushi.oka@keio.jp}{tatsushi.oka@keio.jp})
      } \\
    \end{minipage}
    \\ \\ 
    \begin{minipage}[t]{0.30\textwidth}
      \centering
       Undral Byambadalai\footnote{
       The Economic Institute \& National University of Mongolia 
       (Email: 
       \href{mailto: undral21@gmail.com}{undral21@gmail.com}
       )}\\
    \end{minipage}
    &
    \begin{minipage}[t]{0.30\textwidth}
      \centering
      Yuki Oishi\footnote{
      AbemaTV
      (Email: 
       \href{mailto: oishi_yuki@cyberagent.co.jp}{oishi\_yuki@cyberagent.co.jp}
       )}\\ 
    \end{minipage}
  \end{tabular}
  \vspace{1.0cm}
}

\date{ \normalsize  
\today
}

\begin{document}

\maketitle

\begin{abstract}
\setstretch{1.25}
We examine the impact of top-of-screen promotions on viewing time at ABEMA, a leading video streaming platform in Japan. To this end, we conduct a large-scale randomized controlled trial. Given the non-standard distribution of user viewing times, we estimate distributional treatment effects. Our estimation results document that spotlighting content through these promotions effectively boosts user engagement across diverse content types. Notably, promoting short content proves most effective in that it not only retains users but also motivates them to watch subsequent episodes.

\end{abstract}

\vspace{0.5cm}
{\small 

}

\newpage
\setstretch{1.25}
\section{Introduction}
\label{sec:introduction}

Digital platforms have revolutionized how consumers search for information and make decisions \citep{goldfarb2019digital}. Understanding this search behaviour builds on a rich theoretical tradition, tracing back to seminal works on optimal search under uncertainty \citep{Rothschild1974Searching, Weitzman1979Optimal}. 
Drawing on these insights, modern platforms continuously experiment with various design features to optimize user experience and business outcomes. 
A growing body of research examines how such interventions affect consumer information acquisition through experimental studies, including analyses of search rankings on eBay \citep{dinerstein2018consumer} and advertising disclosure on Zomato \citep{sahni2020does}, among others.
Experiments have also been used to optimize platform treatment rules
\citep[see e.g.][]{ida2022choosing}.

These platform experiments draw on a well-established causal inference literature. Since the foundational works of \cite{fisher1925statistical, fisher1935design} and \cite{Neyman1923}, randomized controlled trials (RCTs) have become the gold standard for evaluating treatment effectiveness. Over the last few decades, RCTs, also known as A/B testing in industry contexts, have emerged as essential tools for causal inference across diverse fields \citep[see e.g.][]{duflo2007using, imbens2015causal}. Their importance is particularly pronounced in digital platform research, where a body of recent empirical work has demonstrated that observational data are often insufficient for establishing causal effects. Studies on platforms such as eBay \citep{blake2015consumer}, Yahoo \citep{lewis2015unfavorable}, and Facebook \citep{gordon2023close}, as well as in television advertising \citep{shapiro2021tv}, have underscored the necessity of experimental methods for credible causal inference.

In this paper, 
we examine how content promotion affects user engagement on streaming platforms by conducting a large-scale RCT on ABEMA, a leading streaming platform in Japan with over 30 million weekly active users.
The platform places promotional banners at the top of the home screen to feature selected content. We randomly assigned users to see either content promotions or advertisements in this position for four weeks and measured total viewing time for the promoted series.

Estimating the treatment effect in our setting poses two main challenges. First, viewing times exhibit both skewness and discreteness. The discreteness occurs because users watch videos of fixed lengths, creating mass points in the distribution. 
Standard average treatment effects (ATE) can obscure important heterogeneity when the outcome follows such a non-normal, mixed discrete-continuous distribution. 
While quantile treatment effects (QTE) have been widely used to capture distributional heterogeneity, most QTE methods require smooth outcome distributions.\footnote{
For continuous outcomes, QTE has been widely used since \citet{doksum1974empirical} and \citet{lehmann1975nonparametrics}, 
inspiring a rich literature on estimation and inference methods \citep[e.g.,][]{heckman1997making, imbens1997estimating, koenker2005quantile, bitler2006mean, athey2006identification, firpo2007efficient, chernozhukov2013inference, koenker2017handbook, belloni2017program, callaway2018quantile, callaway2019quantile}. 
}
Given the discrete nature of our viewing time data, we instead estimate distributional treatment effects (DTE), 
which capture how the treatment shifts the entire outcome distribution and naturally accommodate non-smooth distributions. This approach follows the framework developed by \citet{abadie2002bootstrap}, among others.

The second challenge is low statistical power. Despite their value for causal inference, RCTs on digital platforms often struggle to detect small treatment effects amidst highly variable user behavior \citep{lewis2015unfavorable, johnson2023inferno}. To address this issue, we employ regression adjustment using pre-treatment covariates.
The regression adjustment is a well-established approach for improving precision in ATE estimation
and 
has been extensively studied. See \citet{fisher1925statistical, cochran1977sampling, yang2001efficiency, rosenbaum2002covariance, freedman2008regression, freedman2008regression2, tsiatis2008covariate, rosenblum2010simple, lin2013agnostic, berk2013covariance, ding2019decomposing}, among others.
For estimating distributional treatment effects, we draw on recent methodological developments in regression adjustment within a distributional regression framework for low-dimensional covariates \citep{oka2026regression} and high-dimensional covariates  via machine learning methods \citep{byambadalai24a}. By adapting these approaches, we improve estimation precision and apply theoretically valid inference. 
See also 
\citet{byambadalai2025beyond, byambadalai25a} and \citet{hirata2025efficient}
for complementary extensions of these methods.

Our results demonstrate that promotional campaigns consistently increase initial viewership across all content types, though the persistence of engagement varies significantly depending on content characteristics and narrative structure. Short-form content and programs with strong contextual linkages sustain promotional benefits through continued viewing, while long-form serialized content shows sustained effects only when early episodes provide compelling narrative hooks.  Through distributional and probability treatment effect analyses, we uncover substantial heterogeneous behavioral responses that average treatment effects alone would obscure, revealing not only when promotions drive initial engagement but also the underlying mechanisms that determine viewer retention. These insights enable streaming platforms to strategically target promotions based on content characteristics and inform creators about narrative pacing for improved audience retention.

The paper is organized as follows. We first describe the field experiment setup at ABEMA in Section \ref{sec:experiment}. We then introduce the potential outcomes framework to introduce treatment effect parameters, and explain the estimation method in Section \ref{sec:estimation}. Section \ref{sec:results} presents the empirical results.
Finally, we conclude in Section \ref{sec:conclusion}. The Appendix includes a table detailing covariate balance along with supplementary results.

\section{Experimental Setup at ABEMA}
\label{sec:experiment}

ABEMA is one of the flagship products of CyberAgent, a major technology company in Japan.
Randomized controlled trials (RCTs) have also become an essential component of business practice at CyberAgent. Indeed, hundreds of RCTs are conducted annually on ABEMA. 
These experiments are not used solely to decide whether a specific treatment should be implemented; rather, they serve as an integral learning process through which the firm gains insights into product behavior and consumer responses, thereby informing the development of better products.

Platform engagement within streaming services is fundamentally constrained by the scarcity of user attention.
A prevalent approach for managing this constraint is the differential allocation of content prominence.
For example, ABEMA employs a \emph{top-of-screen promotion} strategy. 
This approach places promotional banners in the 
application's most dominant visual location—the top of the home screen—to feature particular content. 
Given the limited space available, understanding the impact of increased visibility on content consumption is vital. 

\vspace{0.3cm}
\begin{figure}[h]
    \centering
     \caption{ Top-of-Screen Promotion at the ABEMA app} 
\includegraphics[width=0.450\linewidth]{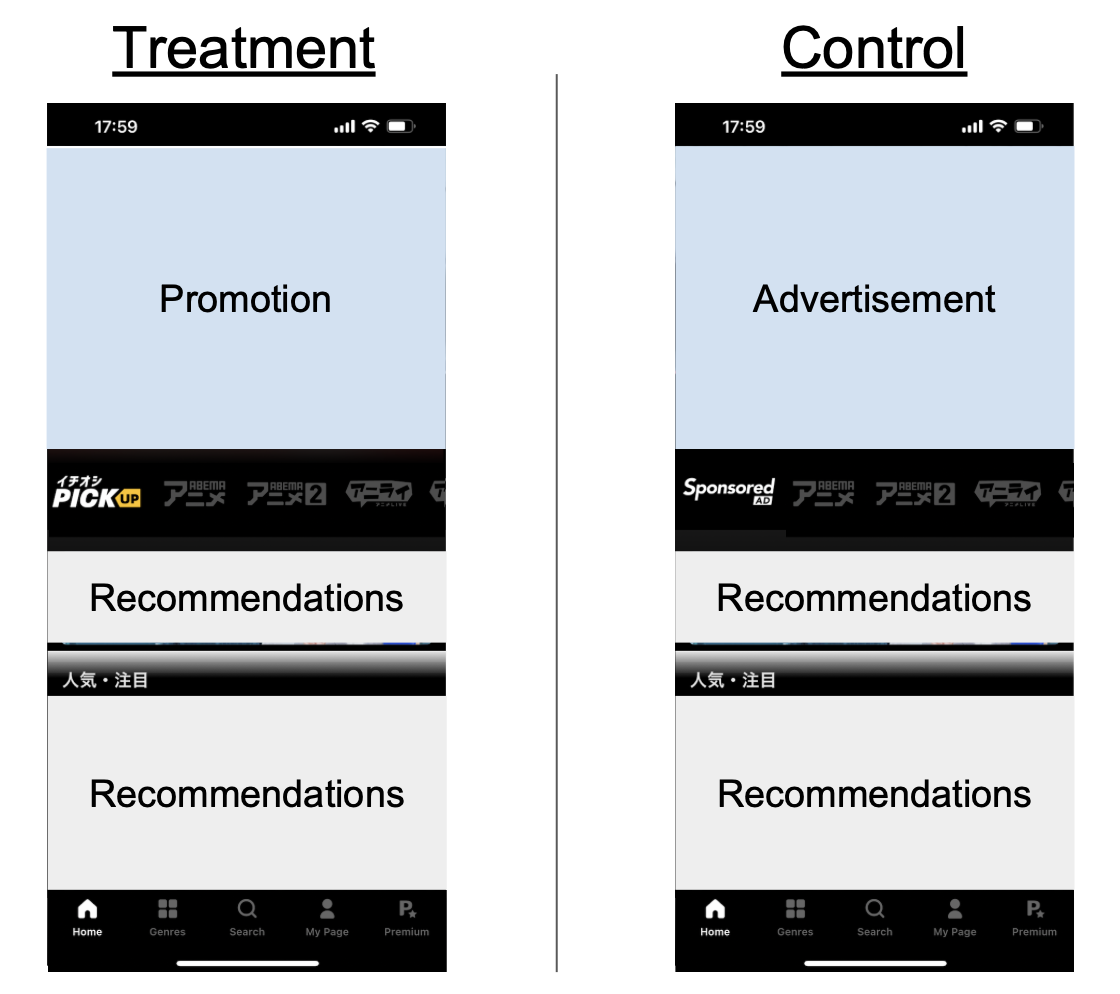}
     \label{fig:APP}
  \begin{minipage}[t]{129mm}
  \begin{spacing}{0.9}  
  {\footnotesize 
  \textit{Notes:} 
  This figure illustrates the ABEMA application display used in the RCT. The left panel (Treatment) presents the application with a content promotion on the top screen. The right panel (Control) presents the application with a standard advertisement in the same location.
  }
  \end{spacing}
  \end{minipage}

\end{figure}
\vspace{0.1cm}

We conducted a large-scale RCT at ABEMA to measure the impact of this strategy, exposing the treatment group to content promotions and the control group to advertisements, 
as illustrated in Figure~\ref{fig:APP}.  
In the experiment, approximately 4.3 million users, which is a subset of the total user base selected at random, were randomly assigned to treatment and control groups.\footnote{Due to confidentiality requirements imposed by ABEMA, the criteria used to determine the size of this subset cannot be disclosed.} 
The treatment group was exposed to content promotions and the control group received advertisements, with a treatment assignment probability of 0.1.
The primary outcome of our RCT is each user's total viewing time of the promoted content.

In this paper, we examine four representative cases that span distinct content types on ABEMA. These include a long-form comedy program, a short sports highlight series, and two serialized reality shows with different narrative structures. This design enables us to compare how the top-of-screen promotion influences engagement across both short and long content as well as between self-contained and continuous formats.

\begin{figure}[h]
    \centering
     \caption{Illustration of Viewing Time Density across Sequential Episodes} 
    \label{figure:view-time}
\begin{tikzpicture}[scale=0.8, transform shape]
\begin{axis}[
    axis lines = middle,
    axis on top,
    xlabel = {Viewing Time},
    ylabel = {},
    ymin=0, ymax=1.2,
    xmin=-1, xmax=7.5,
    xtick={2.5, 5, 6.5}, 
    xticklabels={$t_1$, $t_2$, $t_3$},
    extra x ticks={0},
    extra x tick labels={0},
    extra x tick style={xticklabel style={anchor=north}},
    ytick=\empty,
    axis line style={-stealth, thick},
    every axis y label/.style={at={(axis description cs:0.13,1)}, anchor=south},
    every axis x label/.style={
        at={(axis description cs:0.5,-0.1)}, 
        anchor=north,
        font=\small
    },
    xticklabel style={anchor=north, yshift=-2pt, xshift=2pt},
    clip=false
]

\draw[ultra thick, black] (axis cs:0,0) -- (axis cs:0, 1.1);
\addplot[only marks, mark=*, mark size=2pt, black] coordinates {(0, 1.1)};

\addplot[thick, black, domain=0:2.5, samples=100] 
    {0.70 + (0.57-0.70)*(x/2.5) + 0.06*sin(deg(pi*x/2.5))};

\addplot[thick, black, domain=2.59:5, samples=100] 
    {0.39 + (0.28-0.39)*((x-2.5)/2.5) + 0.01*sin(deg(pi*(x-2.5)/2.5))};

\addplot[thick, black, domain=5.09:6.5, samples=100] 
    {0.10 + (0.05-0.10)*((x-5)/1.5) + 0.01*sin(deg(pi*(x-5)/1.5))};

\draw [thin, dashed] (axis cs:2.5, 0.57) -- (axis cs:2.5, 0.41);
\draw [thin, dashed] (axis cs:5, 0.28) -- (axis cs:5, 0.11);
\draw [thin, dashed] (axis cs:6.5, 0.05) -- (axis cs:6.5, 0);

\addplot[only marks, mark=*, mark size=2pt, black] coordinates {(2.5, 0.57)};
\addplot[only marks, mark=o, mark size=2pt, black, fill=white] coordinates {(2.5, 0.39)};

\addplot[only marks, mark=*, mark size=2pt, black] coordinates {(5, 0.28)};
\addplot[only marks, mark=o, mark size=2pt, black, fill=white] coordinates {(5, 0.10)};

\addplot[only marks, mark=*, mark size=2pt, black] coordinates {(6.5, 0.05)};

\addplot[fill=gray!20, draw=none, domain=0:2.5, samples=100] 
    {0.70 + (0.57-0.70)*(x/2.5) + 0.06*sin(deg(pi*x/2.5))} \closedcycle;

\addplot[fill=gray!20, draw=none, domain=2.5:5, samples=100] 
    {0.39 + (0.28-0.39)*((x-2.5)/2.5) + 0.01*sin(deg(pi*(x-2.5)/2.5))} \closedcycle;

\addplot[fill=gray!20, draw=none, domain=5:6.5, samples=100] 
    {0.10 + (0.05-0.10)*((x-5)/1.5) + 0.01*sin(deg(pi*(x-5)/1.5))} \closedcycle;

\end{axis}
\end{tikzpicture}

  \begin{minipage}[t]{140mm}
  \begin{spacing}{0.9}  
  {\footnotesize 
  \textit{Notes:} 
The figure illustrates the viewing time density across a series of content segments. The horizontal axis represents the viewing time, while the vertical axis depicts the probability density. The mass point at zero viewing time depicts non-engagement. The time markers $t_1$, $t_2$, and $t_3$ denote episode boundaries, where the downward discontinuities (jumps) reflect the discrete probability of users terminating their viewing upon the completion of an episode.
  }
  \end{spacing}
  \end{minipage}

\end{figure}
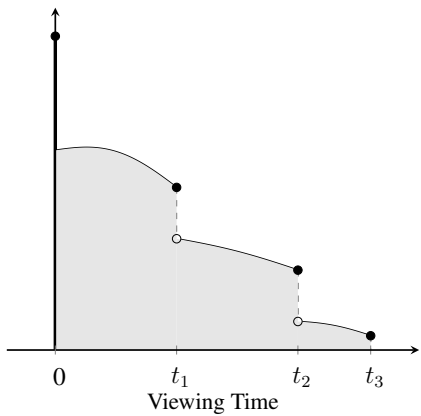
\vspace{0.1cm}

Estimating the treatment effect presents two main challenges. First, viewing times exhibit both skewness and discreteness 
as illustrated in Figure \ref{figure:view-time}. 
The latter characteristic arises from users consuming video content of identical lengths. In such a setting with a non-normal, mixed discrete-continuous distribution, the average treatment effect (ATE) might not adequately reflect the full impact of the treatment. 
Second, RCTs on digital platforms often suffer from low statistical power due to relatively small treatment effects being overshadowed by high variance, as highlighted in the survey by \cite{johnson2023inferno}. 

Experiments on content promotion are frequent on ABEMA. However, given the non-normal distribution mentioned above, standard tools such as the average treatment effect prove inadequate. To obtain more informative results, we adopt an approach that extracts distributional effects with smaller variance, as explained in the next section.

\section{Empirical Framework}
\label{sec:estimation}

In this section, we first define the treatment effect parameters of interest under the potential outcome framework and then describe an estimation method that utilizes pre-treatment covariates to enhance estimation precision.

\subsection{Treatment Effects}

To define treatment effect parameters within a randomized controlled trial (RCT), we employ the potential outcomes framework \citep{rubin1974estimating}.
Let 
$Y(1)$ and $Y(0)$ 
denote the random variables
for
the potential outcomes under treatment and control, respectively. 
Let $D$ be the random variable for treatment assignment indicator, where $D=1$ for treatment and $D=0$ for control.
We denote the cumulative distribution function for 
potential outcome $Y(d)$
as 
$F_{Y(d)}(y):= \Pr(Y(d) \leq y)$ for treatment status $d \in \{0, 1\}$.
The assignment mechanism is a randomized trial with treatment probability $\rho := \Pr(D=1) \in (0,1)$.
We cannot observe $Y(1)$ and $Y(0)$ simultaneously. 
Instead, we observe the scalar outcome $Y$, which satisfies the relation $Y = Y(D)$ with support $\mathcal{Y} \subset \mathbb{R}$, which implicitly assumes the Stable Unit Treatment Value Assumption (SUTVA) or no interference. 

The randomization of $D$ ensures that the treatment assignment is independent of the potential outcomes. This independence, combined with the fact that $0<\rho<1$, allows for the identification of the potential outcome distributions. Therefore, $F_{Y(d)}(y)$ is equivalent to the observable conditional distribution $F_{Y}(y|d):= \Pr(Y\le y|D=d)$
for any $y \in \mathcal{Y}$ and $d \in \{0,1\}$.

The average treatment effect (ATE) is a standard parameter of interest, representing the mean difference in potential outcomes: $\Delta^{ATE}:=  
  \E[ {Y(1)}^{} ] - \E [ Y(0) ]  
$. 
However, as previously discussed, the ATE serves as a measure of central tendency and thus provides limited insight into the treatment's impact on broader distributional features of the outcome variable.
To address the issue, 
we consider the distributional treatment effect (DTE), defined as 
\begin{align*}
  \Delta^{DTE}(y) :=  F_{Y(1)}^{}(y) - F_{Y(0)}^{}(y), \hspace{0.5cm} y \in \R.
\end{align*}
The advantage of the DTE lies in its ability to assess the treatment effect across the entire distribution, while also accommodating various types of outcomes, including mixed discrete-continuous variables.

Moreover, the distributional information is an essential building block for other types of treatment effect parameters. 
For example, the probability  $\Pr(y < Y(d) \le y+h)$ given a constant $h>0$ is equal to the difference $F_{Y(d)}(y+h) - F_{Y(d)}(y)$.
As such, the probability treatment effect (PTE) is defined as 
\begin{align*}
  \Delta^{PTE}(y, h)
  & :=
    \{
    F_{Y(1)}(y+h)
    -
    F_{Y(1)}(y)
    \} -
    \{
    F_{Y(0)}(y+h)
    -
    F_{Y(0)}(y)
    \}, \hspace{0.5cm} y \in \R, h>0.
\end{align*}
The PTE measures changes in the probability that the outcome variable falls in interval $(y, y+h]$ and can be applied for any type of outcome variables.

\subsection{Estimation Method}

Suppose that we obtain a random sample 
$\{(D_{i}, X_{i}, Y_{i})\}_{i=1}^{n}$ of size $n$ from the joint distribution of $(D, X, Y)$. 
Here, $D$ and $Y$ denote the observed treatment assignment and outcome, respectively, 
as defined previously.  
Let  
$X \in \mathcal{X} \subset \mathbb{R}^{p}$ denote a $p$-dimensional vector of pre-treatment covariates,
which are observable characteristics measured prior to treatment assignment. 
The randomization of $D$ ensures that $X$  is independent of the treatment status. 

Under the randomized experiment setup, each unit $i$ 
receives treatment with probability $\rho$, 
yielding $n_1$ treated units and $n_0$ control units
with $n = n_{0} + n_{1}$.
Thus, for the estimation of DTE and PTE, one can use the sample analog estimators simply using the empirical distribution functions of the outcome: 
$n_{1}^{-1} \sum_{i=1}^{n} D_i 
\1_{  \{Y_{i} \le y\} }$
and 
$n_{0}^{-1} \sum_{i=1}^{n} (1-D_i) \1_{  \{Y_{i} \le y\} }$
for $y \in \mathcal{Y}$, where  
$\1_{\{\cdot\}}$ represents the indicator function.

In many experiments, however, we often have access to pre-treatment covariates. 
To incorporate pre-treatment covariates $X$ for improved estimation, we adopt a distribution regression framework. This approach treats the estimation of the conditional distribution function $F_{Y(d)|X}(y|x) := \Pr(Y(d) \le y \mid X=x)$ as a series of binary classification problems indexed by location $y \in \mathcal{Y}$. Specifically, for each $(d, y) \in \{0,1\} \times \mathcal{Y}$, we can express the conditional distribution function as
\begin{align*}
    F_{Y(d)|X}(y|X) = \mathbb{E}[\1_{\{Y(d) \leq y\}} \mid X].
\end{align*}
This formulation enables us to cast the estimation of conditional distribution functions as a supervised classification problem at each location, allowing us to employ a range of statistical and machine learning methods, including parametric approaches (e.g., linear and logistic regression) and flexible nonparametric techniques (e.g., LASSO, random forests, gradient boosting, and neural networks). A key advantage of this framework is its generality: it accommodates continuous, discrete, and mixed-type outcome variables without requiring distributional assumptions.
Let 
$\widehat{F}_{Y(d)|X}(y|x)$ be
an estimator of $F_{Y(d)|X}(y|x)$ based on observations with treatment status $d \in \{0,1\}$  
using any statistical or machine learning method.

In the next step, following \cite{byambadalai24a}, 
we construct a regression-adjusted estimator of the marginal distribution 
$F_{Y(d)}(y)$
defined as, for $d=0,1$,
\begin{align} \label{eq:dte-reg}
    \widehat F_{Y(d)} (y) := & 
    \frac{1}{n_d} \sum_{i: D_i=d} 
    \Big (
    \1_{\{Y_i \leq y\}}-
    \widehat{F}_{Y(d)|X}(y|X_i) 
    \Big ) 
    + \frac{1}{n} \sum_{i=1}^{n} 
    \widehat{F}_{Y(d)|X}(y|X_i).
\end{align}
This estimator takes the familiar doubly-robust form and satisfies a Neyman-orthogonal moment condition \citep{neyman1959optimal, chernozhukov2018debiased}. Neyman orthogonality ensures robustness to first-stage estimation errors by decomposing the estimation problem into nuisance parameter estimation and treatment effect estimation.\footnote{
Our argument builds on recent developments in semiparametric estimation rooted in foundational work by \citet{robinson1988root, bickel1993efficient, newey1994asymptotic, robins1995semiparametric}. 
} 

This formulation is analogous to the regression-adjusted estimator for the average treatment effect,
$\widehat\Delta^{ATE}_{adj}:= \widehat E[Y(1)] - \widehat E[Y(0)]$, 
where, for each $d\in\{0,1\}$, 
\begin{align*} 
    \widehat E[Y(d)] :=& \frac{1}{n_d} \sum_{i: D_i=d} 
    \Big (
    Y_i -
    \widehat{E}[Y(d)|X_i]
    \Big ) 
    + \frac{1}{n} \sum_{i=1}^{n} \widehat {E}[Y(d)|X_i].
\end{align*}
Here, $\widehat{E}[Y(d)|X_i]$ denotes a flexible estimator of the conditional mean function $E[Y(d)|X]$,  
often implemented via machine learning techniques like LASSO or gradient boosting. 
It is based on a doubly-robust moment condition for the ATE under a known propensity score \citep{robins1995semiparametric}. 

In our setting, the nuisance parameters are the conditional outcome distributions $F_{Y(d)|X}(y|x)$. We estimate these distributions using gradient boosting to handle the moderately high-dimensional pre-treatment covariates. 
Theoretical results in 
\cite{byambadalai24a} show that 
we can employ cross-fitting to mitigate the impact of first-stage estimation errors, ensuring valid inference despite the use of flexible machine learning methods.

\section{Empirical Results}  
\label{sec:results}
In the four-week RCT of top-of-screen promotions on ABEMA, approximately 4.3 million users, drawn from the platform’s active user base, were randomly split into two groups: one exposed to content promotions ($D_{i}=1$) and another to advertisements ($D_{i}=0$), with an allocation probability of $\rho = 1/10$. Our dependent variable is the total viewing time during the experimental period for the series to which the promoted content belongs—a series being a set of related episodes. 
When applying the regression adjustment procedures in Equation \eqref{eq:dte-reg}, we use pre-treatment covariates, including 
total viewing time for the same series as the outcome three weeks preceding the experiment, overall content viewing time, gender, and age.
For a complete list of covariates and their balance across treatment groups, see Table \ref{tab:covariates} in Appendix \ref{sec:appendix}. We use gradient boosting (XGBoost) with 3-fold cross-fitting and 500 bootstrap repetitions to compute standard errors and pointwise confidence intervals.

We below present four representative cases (Cases 1-4) that illustrate the typical patterns observed in our experiment. 
Table \ref{tab:ATE} reports both the unadjusted and regression-adjusted estimates of the average treatment effect (ATE). 
The estimation result shows that 
the promotion increases average viewing time by 1.17 to 3.66 seconds (0.0028–0.061 minutes) across cases. Cases 1–3 show statistically significant gains of 6–9\% over control levels. Case 4 yields a larger absolute estimate (0.049 or 0.059 min) but is not significant due to high variability. Regression adjustment slightly tightens standard errors but leaves the overall conclusions unchanged.

\vspace{0.1cm}
\begin{table}[h!]  
\small
   \centering
   \setlength{\tabcolsep}{10pt} 
      \caption{Average Treatment Effect of Content Promotion}  \vspace{-0.4cm} 
    \begin{tabular}{l c c c c} 
    \hline \hline 
            & Case 1 &  Case 2&  Case 3 &  Case 4 \\ \hline
    ATE  &  0.0144 & 0.0028   & 0.061   & 0.059   \\ 
                        & (0.0069) & (0.0011) & (0.0217) & (0.0519)  \\ [5pt]
    ATE (reg.~adjusted) &  0.014  & 0.0028   & 0.059   & 0.0485   \\ 
                                & (0.0056) & (0.0011) & (0.0213) & (0.0471) \\ [5pt]
        Control mean  & 0.244    & 0.0353   & 0.657    & 1.397   \\ 
                            \hline\hline
    \end{tabular} \\
    \begin{minipage}[t]{120mm}
  \begin{spacing}{0.9}  
  {\footnotesize 
  \textit{Notes:} 
    The sample size is 4.3 million. 
    The outcome variable is viewing time in minutes, measured as user engagement with content across Cases 1-4.
  The estimates of 
  the ATE and the regression-adjusted ATE using gradient boosting with 3-fold cross-fitting  are reported 
  with standard errors in parentheses computed from 500 bootstrap replications. 
  The control mean represents average viewing time in the control group.}
  \end{spacing}
  \end{minipage}
    \label{tab:ATE}
    \end{table}

Despite the statistical significance in several cases, the ATE may be a limited metric in this context.
Since 
ABEMA offers a diverse content library tailored to varied user preferences, even popular titles reach only a subset of the overall user base. Consequently, the outcome variable is zero for a large fraction of users
and the ATE becomes considerably small, failing to appropriately capture the behavioral change induced by the promotion.
To address this challenge, we estimate the DTE and PTE. 
For their estimation, we specified a sequence of location values $\{y\}_{j=0}^{J}$ that begins at zero $(y_0 = 0)$ and increases by a constant interval $h$. The interval $h$ was chosen so that the largest location value corresponds to a specific percentile (e.g., the 99th percentile) of the empirical distribution of viewing time, while ensuring that all locations take integer values. This setup provides evenly spaced, interpretable locations that represent the full range of observed viewing behavior. Notably, these locations include specific thresholds such as episode completion, which enables a direct analysis of distinct viewing segments.
We report the estimation results of the DTE and PTE with regression adjustment. The regression adjustment is implemented via gradient boosting with 3-fold cross-fitting. The pointwise confidence intervals are computed using 500 bootstrap replications.

\paragraph{Case 1: Promotion Increases Initial Viewing and First Episode Completion}

\begin{figure}[htbp] \small
    \centering
     \caption{Distributional Treatment Effect and Probability Treatment Effect (Case 1)} 
\includegraphics[width=1.0\linewidth]{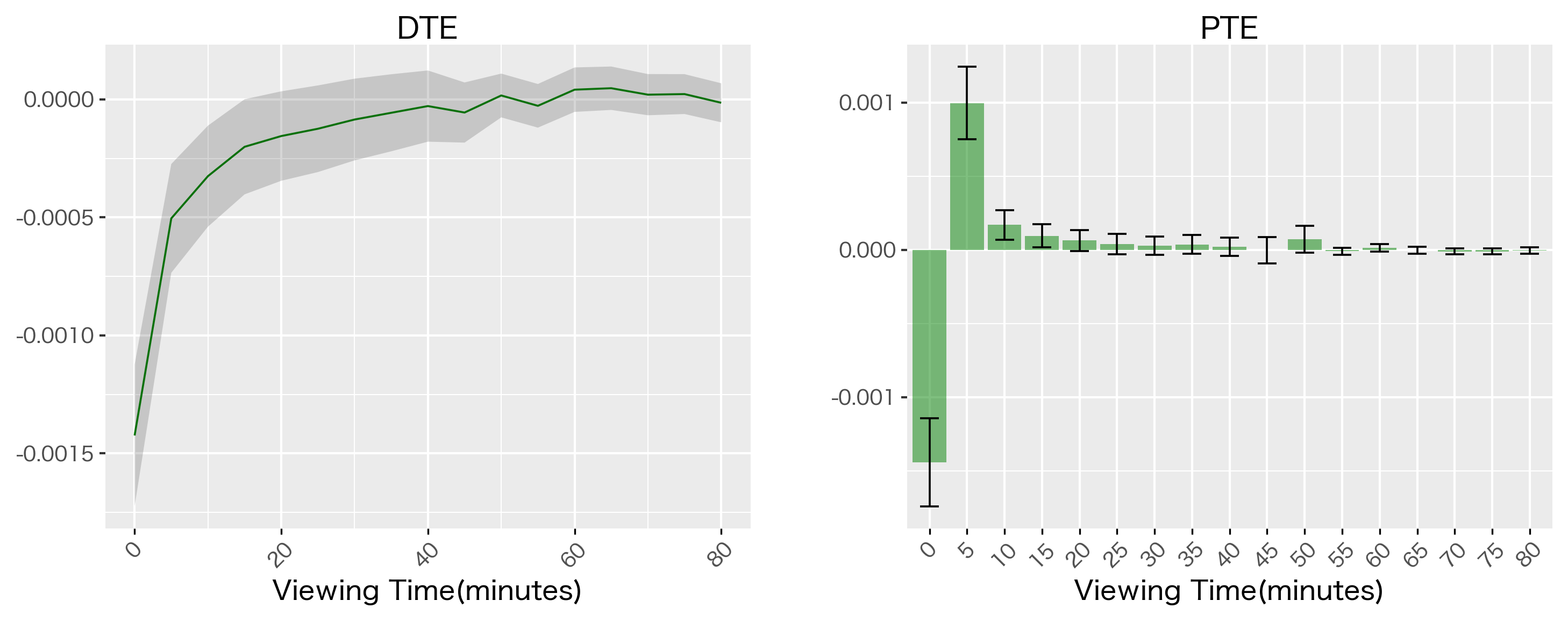}
     \label{fig:DPTE_result1}
         \begin{minipage}{140mm}
           \begin{spacing}{0.8}
  {\footnotesize 
  \textit{Notes:}
  The left panel displays the DTE and the right panel displays the PTE of the content promotion. Both DTE and PTE estimates are obtained using regression adjustment with gradient boosting and 3-fold cross-fitting. The outcome variable is viewing time (minutes). Shaded areas and error bars represent 95\% pointwise confidence intervals based on 500 bootstrap replications.
  }
    \end{spacing}
  \end{minipage}
\end{figure}

Figure \ref{fig:DPTE_result1} presents the estimation results for a 46-minute comedy program. The left panel shows the DTE estimates, where the solid line indicates the DTE and the shaded region outlines the 95\% pointwise confidence interval, whereas the right panel displays the PTE estimation and 95\%  pointwise confidence intervals.

The DTE results reveal that the promotion significantly influences viewing time particularly within the first 20 minutes, after which the estimated effects converge toward zero. 
The PTE results indicate that the promotion decreases the probability of not watching (at 0 minutes) and increases the probability of short-duration viewing, particularly within the first 15 minutes. Beyond these intervals, the estimated effects are small and statistically not significant, indicating that the promotion has little impact on intermediate viewing durations.

In addition, a positive effect is observed around 46 minutes, which corresponds to the end of one episode within the content, although this effect is not statistically significant. This suggests that the promotion may have encouraged a small number of viewers to reach the end of an episode, but its overall influence on sustained viewing remains limited. Moreover, there is no clear tendency for viewers to complete other episodes. Considering that this content consists of self-contained episodes with weak continuity across them, this outcome is a natural consequence. This pattern represents a typical case in which the promotion effectively attracts initial viewers but does not substantially enhance longer-term engagement.

\paragraph{Case 2: Promotion Encourages Multi-Episode Short-Form Viewing}
Secondly, we present the estimation result of a 5-minute sports highlight of particular sport team comprising three matches, with one occurring during the experiment. 

\begin{figure}[htbp] \small
    \centering
     \caption{Distributional Treatment Effect and Probability Treatment Effect (Case 2)} 
\includegraphics[width=1.0\linewidth]{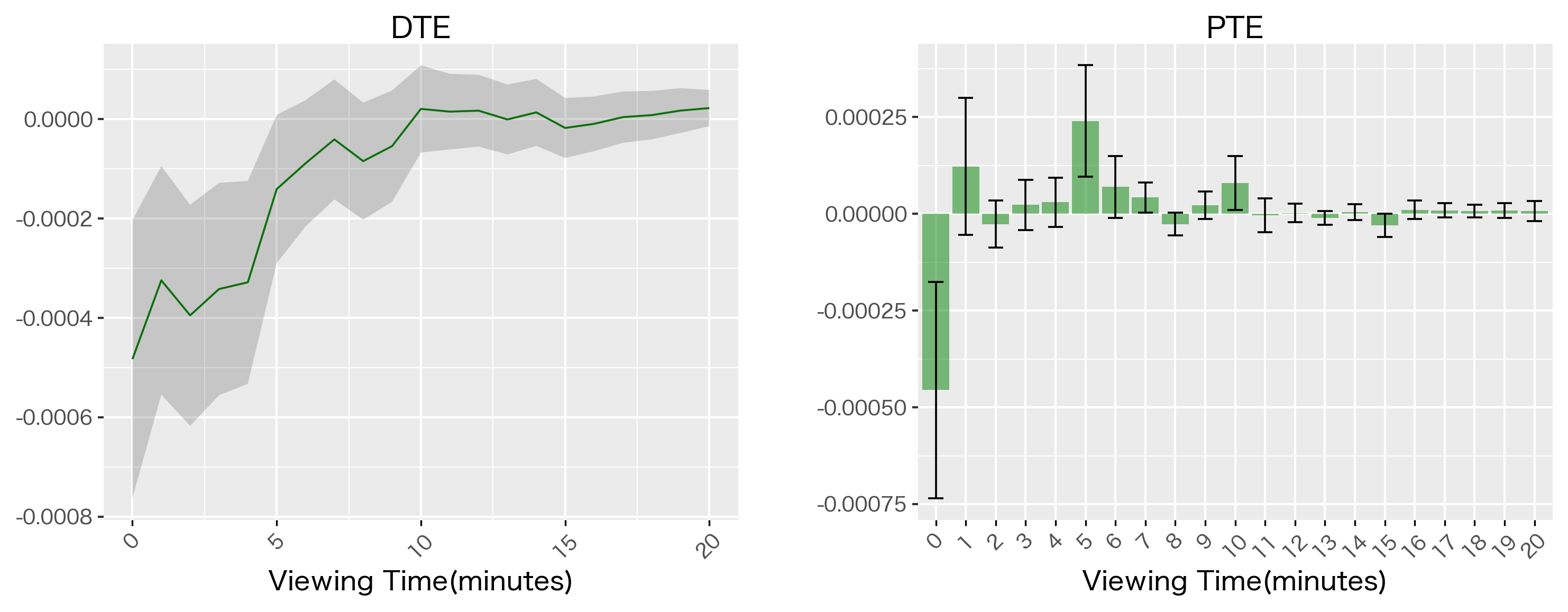}
     \label{fig:DPTE_result2}
         \begin{minipage}{140mm}
           \begin{spacing}{0.8}
  {\footnotesize 
  \textit{Notes:} 
  See the notes to Figure~\ref{fig:DPTE_result1}.
  }
    \end{spacing}
  \end{minipage}
\end{figure}
The left panel of Figure \ref{fig:DPTE_result2} displays the DTE estimation. The results signify that the top-of-screen promotion notably influences users' viewing time within the 0-5 minute range. Given the content length of around 5 minutes, the promotion has increased viewership, with some viewers likely watching a preceding episode.
The PTE results in the right panel of Figure \ref{fig:DPTE_result2} indicate that the promotion decreases the probability of not watching (at 0 minutes) and increases the probability of short-duration viewing, particularly within the first few minutes. Moreover, positive effects are also observed around 5 and 10 minutes, suggesting that viewers tend to complete multiple episodes within the content. This pattern can be attributed to the strong contextual linkage across episodes and the relatively short duration of the content, both of which facilitate consecutive viewing. Beyond these points, the estimated effects are small and statistically not significant, indicating that the promotion’s influence diminishes once viewers have watched several episodes.

\paragraph{Case 3: Promotion Strengthens Long-Form Engagement and Completion}
Next, we present the results for a reality show consisting of eight episodes, each approximately 40 minutes long. This content differs from the previous examples in that the episodes are longer and form a continuous storyline, which allows us to examine how promotional effects unfold in a long-form, serialized context. A key structural characteristic of this program is that the major outcomes are revealed in the final episode, thereby providing a strong incentive for viewers to continue watching once they start.
\begin{figure}[htbp] \small
    \centering
     \caption{Distributional Treatment Effect and Probability Treatment Effect (Case 3)} 
\includegraphics[width=1.0\linewidth]{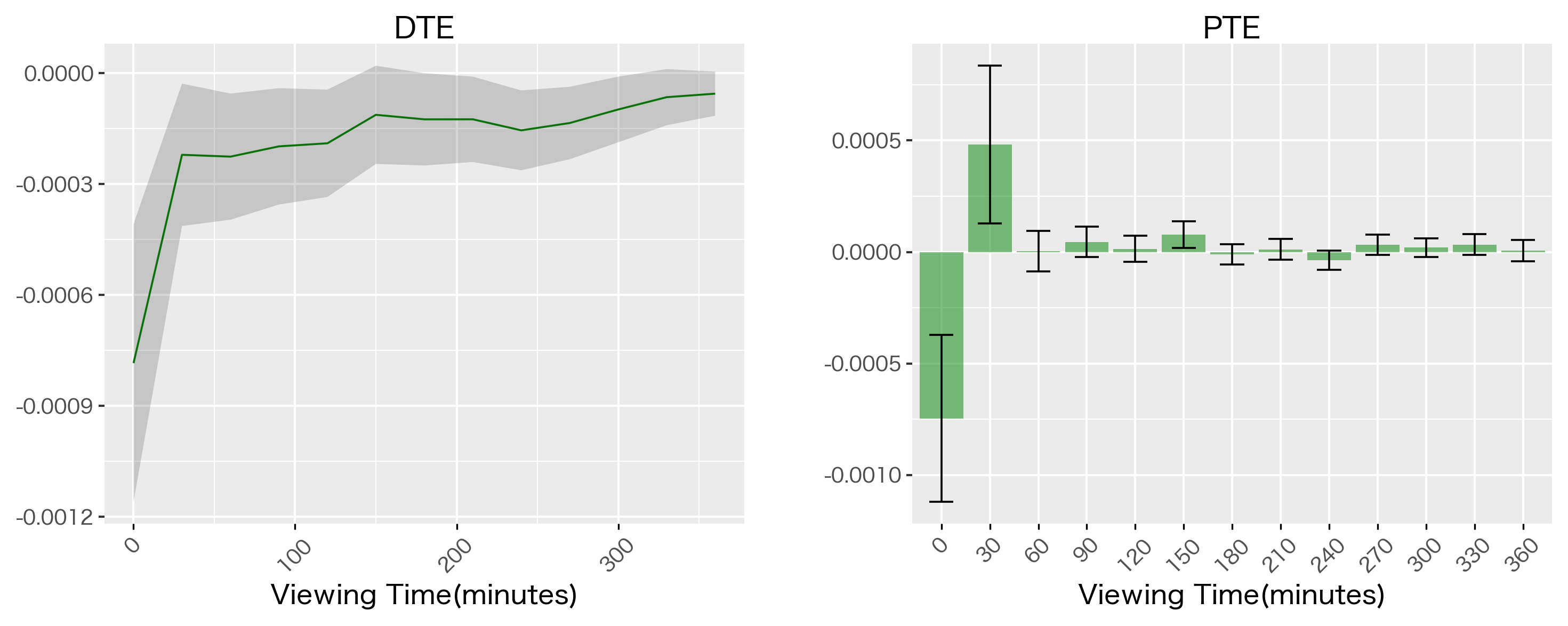}
     \label{fig:DPTE_result3}
         \begin{minipage}{150mm}
           \begin{spacing}{0.8}
  {\footnotesize 
  \textit{Notes:}
  See the notes to Figure~\ref{fig:DPTE_result1}.
  }
    \end{spacing}
  \end{minipage}
\end{figure}
The left panel of Figure \ref{fig:DPTE_result3} displays the DTE estimation. The results indicate that the top-of-screen promotion has a negative effect at the beginning, reflecting the reduction in non-viewers, and the estimated effects gradually increase over time. The DTE remains negative throughout most of the viewing range and becomes more pronounced toward the later part of the content, suggesting that the promotion effectively encourages continued viewing as the series progresses. This pattern is consistent with the structural nature of the reality show, in which major developments occur in the final episodes, providing viewers with a strong incentive to keep watching.

The PTE results in the right panel of Figure \ref{fig:DPTE_result3} show that the promotion decreases the probability of not watching (at 0 minutes) and slightly increases the probability of short-duration viewing, particularly around 30 minutes. In addition, a positive effect is observed around 150 minutes, which likely corresponds to viewers watching multiple episodes consecutively. Moreover, several positive effects are observed between 270 and 330 minutes, suggesting that a small portion of viewers continued to watch through to the final episodes. Although these effects are not statistically significant, they align with the DTE findings, which indicate increased completion rates of this series among promoted users. 

\paragraph{Case 4: Promotion Induces Only Trial Viewing Without Sustained Engagement}
The previous reality show demonstrated that the promotion effectively encouraged continued viewing throughout the series, consistent with its structural design in which major developments occur in the final episodes.
\begin{figure}[htbp] \small
    \centering
     \caption{Distributional Treatment Effect and Probability Treatment Effect (Case 4)} 
\includegraphics[width=1.0\linewidth]{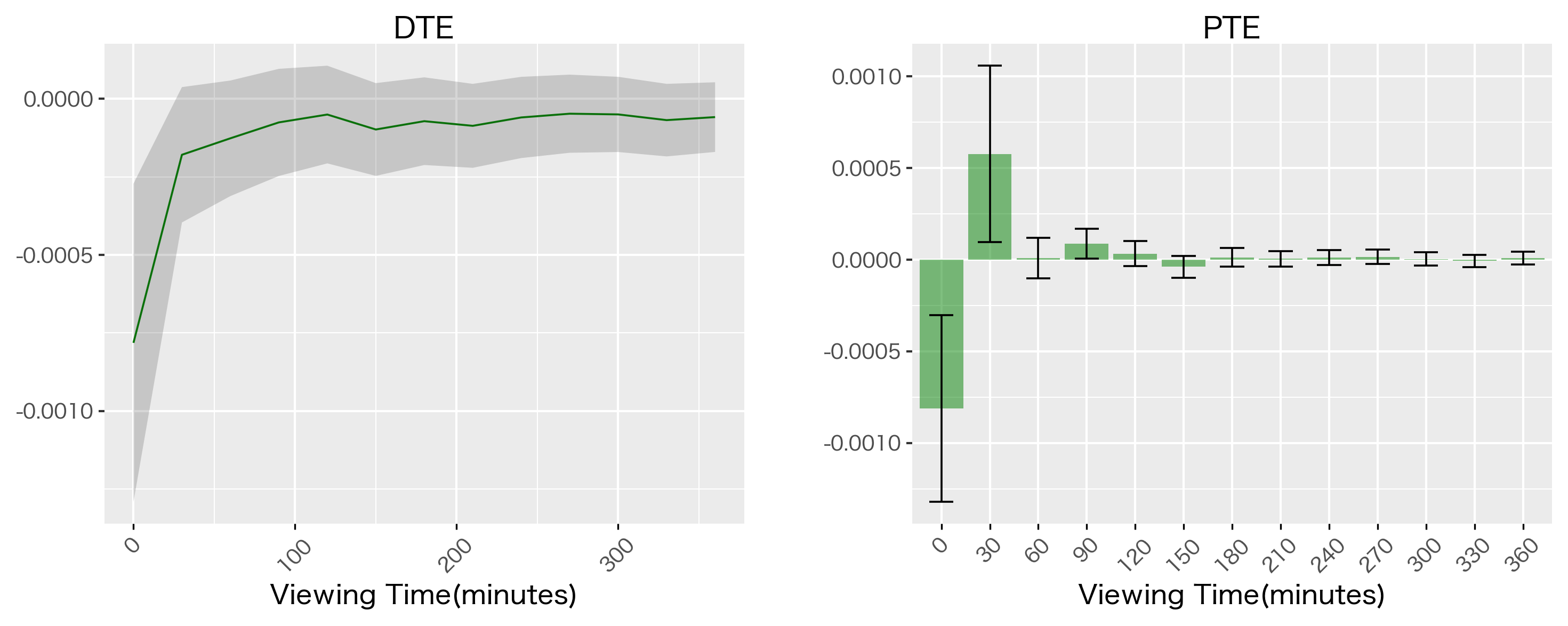}
     \label{fig:DPTE_result4}
         \begin{minipage}{140mm}
           \begin{spacing}{0.8}
  {\footnotesize 
  \textit{Notes:}
  See the notes to Figure~\ref{fig:DPTE_result1}.
  }
    \end{spacing}
  \end{minipage}
\end{figure}
In contrast, Figure \ref{fig:DPTE_result4} presents the results for another reality show with a similar format but a different outcome. The DTE estimates on the left panel indicate that the effects are slightly negative at the beginning and remain close to zero afterward, suggesting that the promotion has little impact on overall viewing duration. The PTE results on the right panel show that the promotion decreases the probability of not watching (at 0 minutes) and slightly increases the probability of short-duration viewing at the start. In addition, a small positive but not statistically significant effect is observed around 90 minutes, which may correspond to viewers watching a second episode. These patterns imply that the promotion successfully motivated users to start watching—most likely the first episode—but that many viewers discontinued thereafter, resulting in no significant increase in long-term viewing or series completion.

Unlike the previous case, the ATE is not statistically significant as shown in Table \ref{tab:ATE}, indicating that the average treatment effect alone fails to capture the short-term behavioral responses identified through the DTE and PTE analyses. This example highlights the usefulness of decomposing the ATE into distributional and probability treatment effects, which reveal transient engagement patterns that are otherwise hidden in the average result. Moreover, the absence of continued viewing suggests potential issues in the early part of the content—such as insufficient engagement or weak narrative development—that limit viewers’ willingness to proceed beyond the first episode.

\paragraph{Key Findings across Content Types}
Table \ref{tab:case_summary} provides an overview of the estimation results across Cases 1-4, highlighting how variations in content format and narrative continuity lead to heterogeneous promotional impacts. In short-form and episodic contents, promotions mainly increase short-duration viewing and encourage users to sample the content, while in long-form serialized programs, the effects extend to sustained engagement when the storyline provides a strong incentive to continue. Conversely, when the initial episodes lack appeal, the promotional effects diminish quickly, as observed in the second reality show.

\begin{table}[htbp]
\centering
\caption{Summary of Viewing Responses to Promotion Across Content Types}
\label{tab:case_summary}
\renewcommand{\arraystretch}{1.0} 
\begin{tabularx}{\textwidth}{c >{\RaggedRight}p{3.5cm} >{\RaggedRight}p{3.5cm} >{\RaggedRight}X}
\hline \hline 
\textbf{Case} & \textbf{Content Type} & \textbf{Episode Structure} & \textbf{Summary of Treatment Effect Pattern} \\
\midrule
1 & Comedy program & Self-contained episodes (46 min) & Promotion increases initial viewing but engagement does not extend to additional episodes due to weak cross-episode continuity. \\ \addlinespace[8pt]
2 & Sports Highlight & Short, sequential clips (5 min) & Promotion increases starting rates and leads to multi-episode viewing, supported by short duration and strong contextual linkage across clips. \\ \addlinespace[8pt]
3 & Reality TV Show A & Long-form serialized narrative with major payoff in final episodes & Promotion increases initial viewing, and the narrative structure encourages continued watching through later episodes, resulting in sustained engagement. \\ \addlinespace[8pt]
4 & Reality TV Show B & Long-form serialized narrative with weak early hook & Promotion increases the chance of trial viewing only; viewers tend to drop after the first episode, yielding no meaningful continuation or completion. \\
\bottomrule
\end{tabularx}
\end{table}

These findings highlight the importance of considering content characteristics when designing and evaluating promotional strategies. Moreover, the DTE and PTE estimations offer valuable insights not only for evaluating promotional effectiveness but also for informing content development. By revealing when and how viewers engage or disengage, these analyses can guide creators in improving narrative pacing and audience retention. From a managerial perspective, understanding these patterns also helps platforms select which types of content are most suitable for promotion, enabling more efficient allocation of promotional resources.

Finally, the comparison between ATE, DTE, and PTE results underscores the analytical advantage of decomposing treatment effects, as it allows us to detect heterogeneous behavioral responses that are otherwise hidden in the average effect. Together, these analyses provide a  comprehensive understanding of how contents promotions influence user engagement in streaming platforms.

\paragraph{Observed Heterogeneity}
We examine treatment effect heterogeneity along gender, motivated by the nature of the media content analyzed. Audience engagement with specific types of content often differs systematically by gender, making it a plausible dimension along which treatment effects may vary. In our setting, the titles in Cases 1 and 2 are more popular among male users, whereas those in Cases 3 and 4 are more popular among female users. Consistent with these baseline popularity patterns, we observe systematic differences in the conditional distributional treatment effects across genders.

\begin{figure}[htbp] 
    \small
    \centering
     \caption{Conditional Distributional Treatment Effect and Probability Treatment Effect} 
\includegraphics[width=1.0\linewidth]{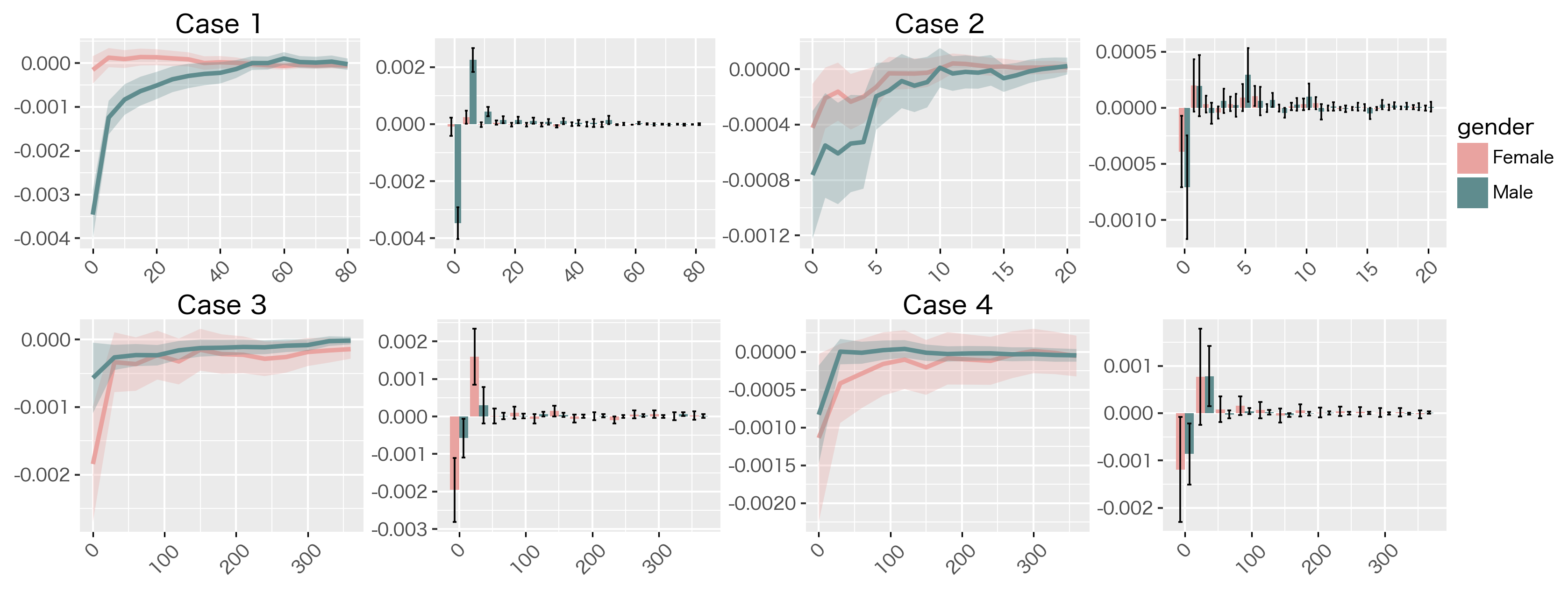}
     \label{fig:DPTE_conditional}
         \begin{minipage}{140mm}
           \begin{spacing}{0.8}
  {\footnotesize 
  \textit{Notes:}
  Estimation results for male and female users are shown in blue and red, respectively. 
  See also the notes to Figure~\ref{fig:DPTE_result1}.
  }
    \end{spacing}
  \end{minipage}
  \label{fig:gender}
\end{figure}

We estimate gender-conditional DTEs and PTEs by applying the same framework separately to female and male users. Figure \ref{fig:gender} presents the results for gender-conditional DTEs and PTEs for Cases 1-4.

For Cases 1 and 2, the treatment effects conditional on being male are substantially larger in magnitude than those conditional on being female. In contrast, for Cases 3 and 4, the conditional effects for female users exceed those for male users, indicating a reversal in the relative magnitude of the effects across genders. 
For Case 2, episodes are sufficiently short that most users attracted by the treatment end up completing the episode. As a result, the DTE exhibits a distinctive step-like pattern. Although the size of the effect differs by gender, this completion-driven dynamic is present for both male and female users, leading to similar temporal patterns of the DTEs across genders.

\section{Conclusion}
\label{sec:conclusion}

In this paper, we estimated the effects of content promotions on user viewing time through a large-scale randomized controlled trial conducted on ABEMA.
Rather than focusing solely on average effects, we analyzed the treatment's impact across the entire viewing time distribution and found that content promotion primarily increases the number of users with short viewing durations. Our results demonstrate that promoting shorter content formats yields greater benefits, as evidenced by increased multi-content consumption when such promotional strategies are implemented.

While this study focuses on randomized content promotion, future research could explore how personalized targeting mechanisms influence user behavior and content discovery on digital platforms. Moreover, extending the analysis of distributional treatment effects from a statistical decision-theoretic perspective represents a promising avenue for future work.

\clearpage
\setlength{\bibsep}{3pt}  
\bibliographystyle{apalike}
\bibliography{ATE_RA.bib}


\newpage
\appendix

\pagenumbering{arabic} 
\renewcommand{\thepage}{A-\arabic{page}} 
\renewcommand{\thepage}{A-\arabic{page}} 
\renewcommand{\thetable}{A\arabic{table}} 
\renewcommand{\thefigure}{A\arabic{figure}} 
\setcounter{table}{0} 
\setcounter{figure}{0} 

\section{Appendix}
\label{sec:appendix}

In this appendix, we first provide information on pre-treatment covariates. We then present estimation results for four additional cases that supplement the main findings.

\subsection{Pre-treatment Covariates}
In the regression adjustment for estimating the ATE, DTE, and PTE, we included a common set of pre-treatment covariates. The list of covariates and their balances are reported in Table~\ref{tab:covariates}. For each case (Cases 1–8), a content-specific covariate was included, representing the total viewing time of the target series during the three weeks prior to the experiment. User attributes consisted of gender, a binary indicator for new users who had not used the service before the experiment, and age group dummies covering users in their teens through sixties (age10s–60s+). We also included the number of days elapsed since a user first registered for the service at the beginning of the experiment (User duration) as a proxy for overall service experience. In addition, we controlled for users’ general viewing activity using the total viewing time of all content during the three weeks prior to the experiment (Prior elapsed time). Finally, we incorporated thirteen genre preference dummies (Preferred content category) based on users’ self-reported preferred content categories at registration. Together, these covariates capture both demographic and behavioral characteristics of users, allowing for a more accurate comparison between the treatment and control groups.

\begin{table}\centering
\caption{Summary Statisics of Pre-Treatment Covariates}
\vspace{-0.2cm}
\label{tab:covariates}
\begin{tabular}{lccc}
\toprule
Variable & Mean Diff & Standard Error & $t$-stat. \\
\midrule
Total viewing time pre-experiment (Case1) & 0.0003 & 0.0052 & 0.0669 \\
Total viewing time pre-experiment (Case2) & 0.0005 & 0.0014 & 0.3326 \\
Total viewing time pre-experiment (Case3) & -0.0012 & 0.0017 & -0.7176 \\
Total viewing time pre-experiment (Case4) & -0.0147 & 0.0293 & -0.5017 \\
Total viewing time pre-experiment (Case5) & -0.0051 & 0.0034 & -1.5267 \\
Total viewing time pre-experiment (Case6) & 0.0080 & 0.0076 & 1.0562 \\
Total viewing time pre-experiment (Case7) & -0.0009 & 0.0016 & -0.5734 \\
Total viewing time pre-experiment (Case8) & 0.0161 & 0.0227 & 0.7104 \\
Female & -0.0003 & 0.0008 & -0.4462 \\
Male & 0.0001 & 0.0008 & 0.1591 \\
New user & 0.0001 & 0.0003 & 0.2709 \\

Age group 10s & -0.0005 & 0.0006 & -0.8081 \\
Age group 20s & 0.0001 & 0.0007 & 0.2072 \\
Age group 30s & 0.0001 & 0.0006 & 0.2091 \\
Age group 40s & -0.0007 & 0.0006 & -1.2168 \\
Age group 50s & 0.0002 & 0.0005 & 0.4302 \\
Age group 60s+ & 0.0010 & 0.0004 & 2.6922 \\
User duration & -0.2835 & 1.1205 & -0.2530 \\
Prior elapsed time & -0.3639 & 0.8251 & -0.4410 \\
Preferred content category (genre 1) & -0.0004 & 0.0005 & -0.8333 \\
Preferred content category (genre 2) & -0.0001 & 0.0004 & -0.2050 \\
Preferred content category (genre 3) & -0.0002 & 0.0002 & -0.7770 \\
Preferred content category (genre 4) & -0.0002 & 0.0003 & -0.5263 \\
Preferred content category (genre 5) & -0.0001 & 0.0004 & -0.4130 \\
Preferred content category (genre 6) & -0.0002 & 0.0003 & -0.6451 \\
Preferred content category (genre 7) & 0.0000 & 0.0003 & 0.1436 \\
Preferred content category (genre 8) & -0.0002 & 0.0004 & -0.4877 \\
Preferred content category (genre 9) & -0.0002 & 0.0001 & -1.3883 \\
Preferred content category (genre 10) & 0.0002 & 0.0001 & 1.6120 \\
Preferred content category (genre 11) & 0.0001 & 0.0001 & 0.4436 \\
Preferred content category (genre 12) & -0.0002 & 0.0002 & -0.7940 \\
Preferred content category (genre 13) & -0.0000 & 0.0001 & -0.7462 \\
\bottomrule
\end{tabular}
 \begin{minipage}[t]{150mm}
  \begin{spacing}{0.9}  
  {\footnotesize 
  \textit{Notes:} 
    This table presents the covariate balance between the treatment and control groups based on a sample of approximately 4.3 million users. The ``Mean Diff'' column reports the difference in covariate means between the treatment group ($D=1$) and the control group ($D=0$). The second and third columns display the corresponding standard errors and t-statistics, respectively, for each mean difference. This balance check helps assess the effectiveness of randomization and the comparability of groups prior to treatment assignment.
    }
  \end{spacing}
  \end{minipage}
\end{table}

\clearpage
\subsection{Other Cases}

This subsection introduces additional four cases (Cases 5–8), offering results that complement the main findings.
For each case, we report the estimated Average Treatment Effect (ATE) along with the Distributional Treatment Effect (DTE) and Probability Treatment Effect (PTE) results. 

Table \ref{tab:results_appendix} reports the unadjuted and regression-adjusted ATE.
Case 5 shows no statistically significant ATE, and both DTE and PTE estimates indicate negligible promotional influence on short-form serialized content.
In contrast, Case 6 exhibits a statistically significant ATE of approximately 8\%, with the DTE and PTE suggesting that the promotion modestly increased viewing across multiple episodes of a long news program.
Case 7 demonstrates a larger and significant ATE of about 20\%, but the DTE and PTE results reveal that the promotional effect was concentrated in attracting initial viewers rather than sustaining longer engagement.
Finally, Case 8 shows a significant ATE of around 16\%, yet neither DTE nor PTE indicates extended viewing, implying that the effect did not translate into longer-form engagement. Together, these results underscore that the top-of-screen promotion’s effectiveness varied substantially by content type and viewing context.

\begin{table}[htbp] \centering
  \setlength{\tabcolsep}{10pt} 
      \caption{Average Treatment Effect of Content Promotion}  \vspace{-0.4cm} 
    \begin{tabular}{l c c c c } 
    \hline \hline 
                        & Case 5 & Case 6& Case 7  & Case 8\\ \hline
                    ATE  & 0.007   & 0.0136  & 0.0093   & 0.0796 \\ 
                        &  (0.0044) & (0.0091)& (0.0033) & (0.0355) \\ \addlinespace[5pt]
    ATE (reg.~adjusted)  &  0.0015  & 0.0201  &  0.0092 &   0.0894 \\ 
                        &  (0.0028) & (0.0068) & (0.0032) & (0.0338) \\\addlinespace[5pt]
            Control mean  &  0.102    & 0.250   & 0.044    & 0.497 \\ \hline    
    \end{tabular} 
    \begin{minipage}[t]{120mm}
  \begin{spacing}{0.9}  
  {\footnotesize 
  \textit{Notes:} 
  See the notes to Table~\ref{tab:ATE}.
  }
  \end{spacing}
  \end{minipage}
\label{tab:results_appendix}
\end{table}

\begin{figure}[htbp] \small
    \centering
     \caption{Distributional Treatment Effect and Probability Treatment Effect (Case 5)} 
\includegraphics[width=1.0\linewidth]{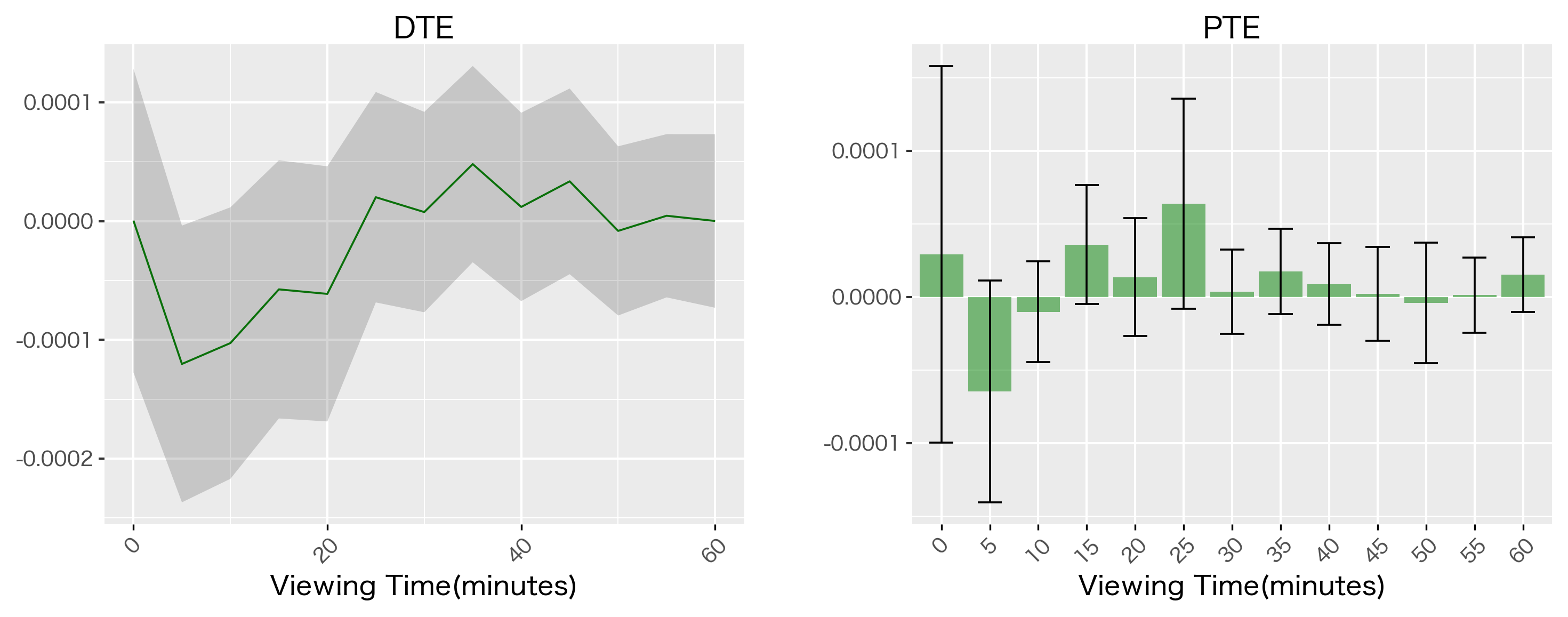}
     \label{fig:DPTE_result5}
         \begin{minipage}{150mm}
           \begin{spacing}{0.8}
  {\footnotesize 
  \textit{Notes:}
  The left panel displays the DTE and the right panel displays the PTE of the content promotion. The outcome variable is viewing time (minutes). Shaded areas and error bars represent 95\% pointwise confidence intervals.
  }
    \end{spacing}
  \end{minipage}
\end{figure}

\paragraph{Case 5: Promotion Shows Negligible Effects on Short-Form Content}
Figure \ref{fig:DPTE_result5} presents the DTE and PTE estimations for Case 5, a short serialized program with an episode length of approximately 23 minutes.

The left panel shows that the DTE fluctuates slightly around zero throughout the viewing range, with a small negative value observed in the early period but no statistically significant effects overall. This indicates that the top-of-screen promotion had little influence on total viewing time.

The right panel displays the PTE estimation with regression adjustment and 95\% confidence intervals. The PTE results reveal minor variations across viewing durations, including a small positive value around 25 minutes—roughly corresponding to the end of one episode—but none of these effects are statistically significant. These results suggest that the promotion neither meaningfully increased the likelihood of starting the content nor affected subsequent viewing behavior. Overall, the DTE and PTE analyses consistently indicate that the promotional intervention produced negligible effects for this content.

\begin{figure}[htbp] \small
    \centering
     \caption{Distributional Treatment Effect and Probability Treatment Effect (Case 6)} 
\includegraphics[width=1.0\linewidth]{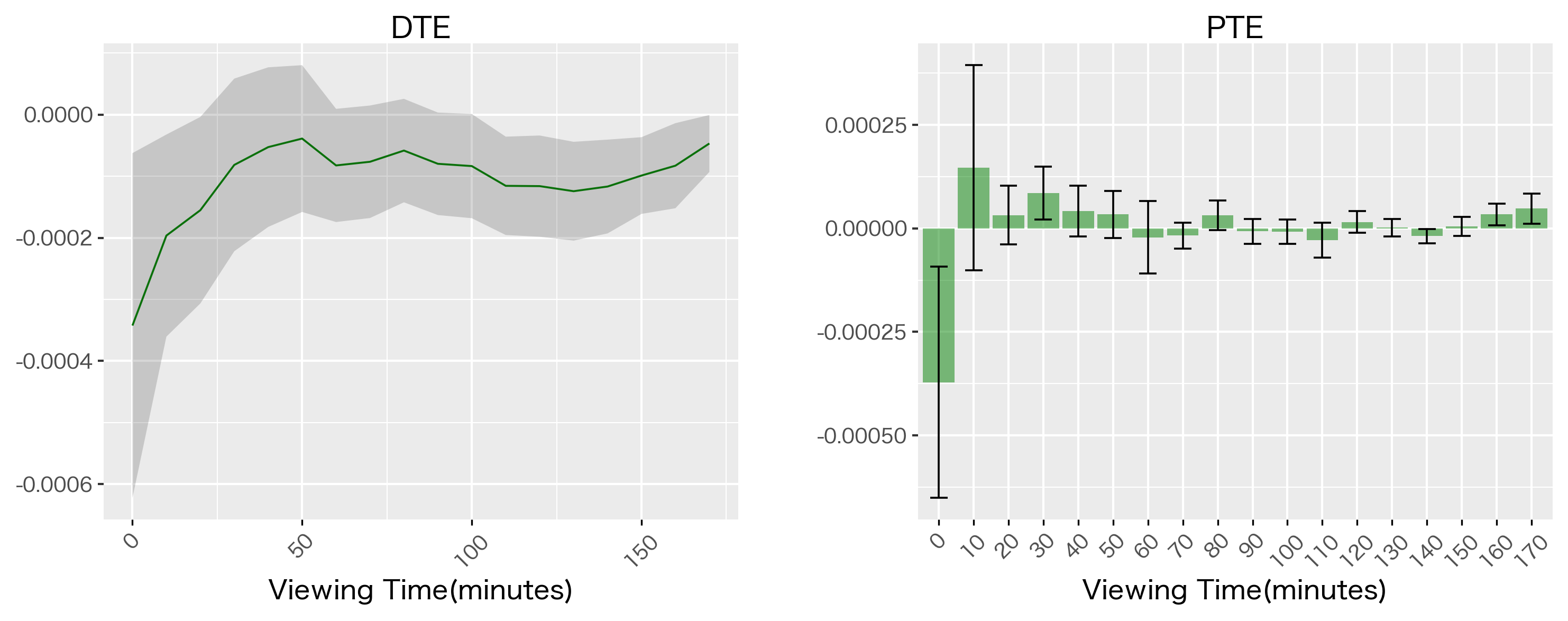}
     \label{fig:DPTE_result6}
         \begin{minipage}{150mm}
           \begin{spacing}{0.8}
  {\footnotesize 
  \textit{Notes:}
  See the notes to Figure~\ref{fig:DPTE_result5}.
  }
    \end{spacing}
  \end{minipage}
\end{figure}

\paragraph{Case 6: Minimal Promotional Influence on a Long News Program}
Figure \ref{fig:DPTE_result6} presents the DTE and PTE estimations for Case 6, a 60-minute news program consisting of multiple informational segments.

The left panel shows that the DTE values are slightly negative in the early viewing period and gradually increase over time. A statistically significant positive effect is observed between 100 and 150 minutes, indicating that the promotion encouraged longer viewing durations among a subset of users. This result suggests that, while the initial response to the promotion was modest, it contributed to sustained engagement later in the viewing period.
The PTE results similarly show a positive effect in 160 and 170 minutes, consistent with the DTE findings. This pattern implies that the promotion increased the probability of users continuing to watch through to the later segments of the program. Beyond this interval, the estimated effects are small and statistically insignificant. Taken together, the DTE and PTE results indicate that the top-of-screen promotion modestly enhanced long-form engagement for this content.

\begin{figure}[htbp] \small
    \centering
     \caption{Distributional Treatment Effect and Probability Treatment Effect (Case 7)} 
\includegraphics[width=1.0\linewidth]{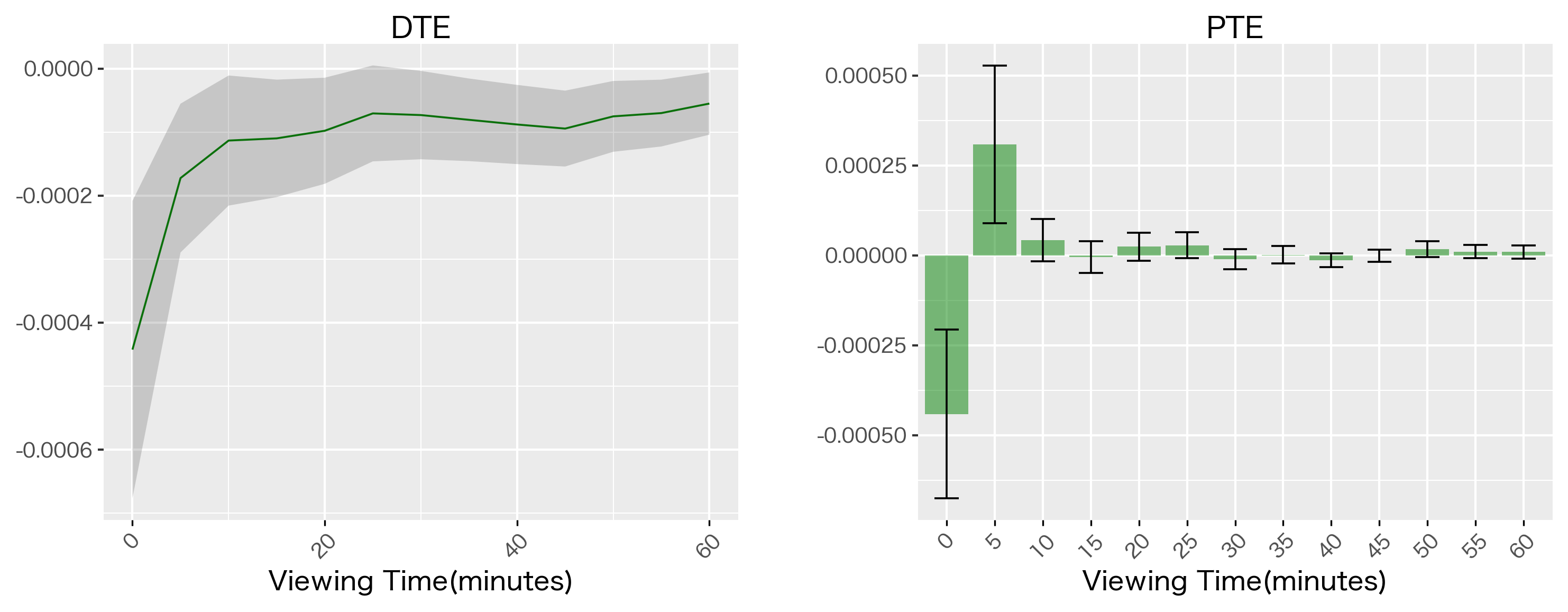}
     \label{fig:DPTE_result7}
         \begin{minipage}{150mm}
           \begin{spacing}{0.8}
  {\footnotesize 
  \textit{Notes:}
    See the notes to Figure~\ref{fig:DPTE_result5}.
  }
    \end{spacing}
  \end{minipage}
\end{figure}

\paragraph{Case 7: Promotion Attracts Initial Viewers}
Figure \ref{fig:DPTE_result7} presents the DTE and PTE estimations for Case 7, a 50-minute stand-alone comedy program.

The left panel shows that the DTE values are significantly different from zero across most of the viewing range. The effects are negative at the very beginning, reflecting a reduction in non-viewers, and then increase and remain positive for the majority of the duration. This pattern indicates that the top-of-screen promotion had a consistent and statistically significant impact on extending viewing time throughout the program.

The PTE results show that the promotion decreases the probability of not watching (at 0 minutes) and increases the probability of short-duration viewing in the early part of the program. Although the estimated effects diminish over time, they remain slightly positive throughout the viewing range, consistent with the DTE findings.

Taken together, the DTE and PTE results suggest that the promotion successfully attracted some users to start watching but did not substantially increase sustained viewing or completion rates. Considering that this program consists of a single, self-contained episode, such short-term engagement without extended viewing is consistent with its stand-alone nature.

\begin{figure}[htbp] \small
    \centering
     \caption{Distributional Treatment Effect and Probability Treatment Effect (Case 8)} 
\includegraphics[width=1.0\linewidth]{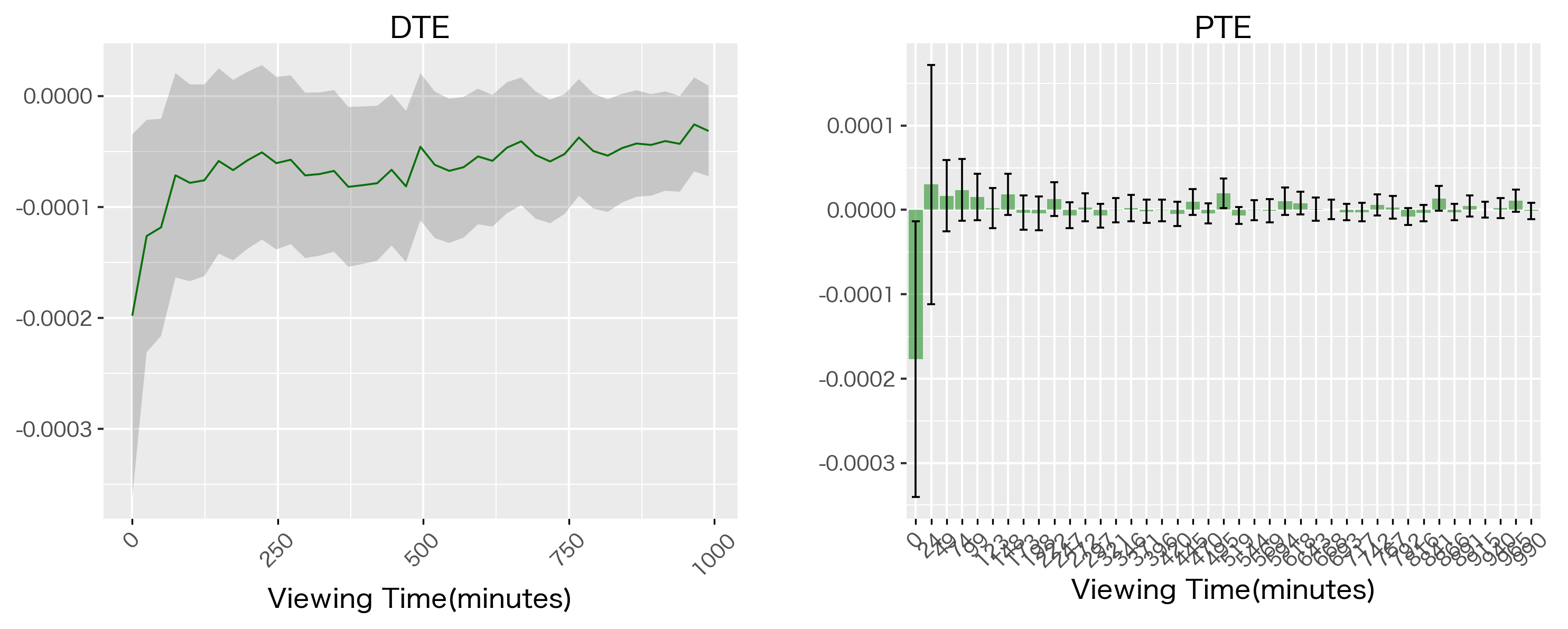}
     \label{fig:DPTE_result8}
         \begin{minipage}{150mm}
           \begin{spacing}{0.8}
  {\footnotesize 
  \textit{Notes:}
    See the notes to Figure~\ref{fig:DPTE_result5}.
  }
    \end{spacing}
  \end{minipage}
\end{figure}

\paragraph{Case 8: No Detectable Promotional Effect on Long-Form Serialized Content.}
Figure \ref{fig:DPTE_result8} presents the DTE and PTE estimations for Case 8, a long-form serialized program exceeding 900 minutes in total viewing time.

The left panel shows that the DTE values are slightly negative at the beginning and gradually approach zero, with wide confidence intervals that overlap the baseline throughout the range. This pattern indicates that the top-of-screen promotion had little to no measurable impact on the overall distribution of viewing time. The PTE result on the right panel shows small positive and negative fluctuations, but none are statistically significant. These findings suggest that the promotion neither increased the likelihood of starting the program nor affected viewing continuity or completion rates.

Taken together, the DTE and PTE analyses provide no evidence of a meaningful promotional effect for this content. Given the long-form serialized structure, the absence of a detectable effect may reflect that users’ viewing behavior is primarily driven by pre-existing interest in the series rather than promotional exposure.

\vspace{1cm}

\end{document}